\documentclass[a4paper,12pt,twoside]{article}
\usepackage{amsmath,amsfonts,amsthm,graphicx}
\numberwithin{equation}{section} 
\newtheorem{theorem}{Theorem}[section]
\newtheorem{conjecture}[theorem]{Conjecture}
\begin{document}
\title{Nonlinear integral equations for thermodynamics of the 
$sl(r+1)$ Uimin-Sutherland model}
\author{ 
Zengo Tsuboi
\footnote{E-mail address: ztsuboi@poisson.ms.u-tokyo.ac.jp}
\\
{\it Graduate School of Mathematical Sciences, 
University of Tokyo,} \\ 
{\it Komaba 3-8-1, Meguro-ku, Tokyo 153-8914, Japan}
}
\date{}
\maketitle
\begin{abstract}
We derive 
traditional thermodynamic Bethe ansatz (TBA) 
 equations for the $sl(r+1)$ Uimin-Sutherland model 
 from the $T$-system of the quantum transfer matrix.
  These TBA equations 
 are identical to the ones from the string hypothesis. 
Next we derive a new family of 
nonlinear integral equations (NLIE). 
In particular, a subset of these NLIE forms a 
system of NLIE which contains only a {\em finite} number of 
 unknown functions. 
For $r=1$, 
this subset of NLIE reduces to 
Takahashi's NLIE for the $XXX$ spin chain. 
A relation between the traditional TBA equations and 
our new NLIE is clarified. 
Based on our new NLIE, we also calculate 
the high temperature expansion of the free energy. 
\end{abstract}
Short title: Nonlinear integral equations \\
{\it PACS2001:} 02.30.Rz; 02.30.Ik; 05.50.+q; 05.70.-a \\
{\it MSC:} 82B23; 45G15; 82B20; 17B80 \\
{\it Key words:}
nonlinear integral equation; 
Uimin-Sutherland model; 
quantum transfer matrix; 
solvable lattice model; 
thermodynamic Bethe ansatz; 
$T$-system \\
{\bf to appear in J. Phys. A: Math. Gen.}\\\\
%
\section{Introduction}
Thermodynamic Bethe ansatz (TBA) equations have 
been used to investigate 
thermodynamics of various kind of solvable lattice models 
(see for example \cite{Ta99}). 
Traditionally, TBA equations have been derived by the 
string-hyposesis \cite{T71,G71}. 
Several years ago, TBA equations for 
the supersymmetric $t-J$ model and 
the supersymmetric extended Hubbard model were 
derived \cite{JKS98} from the $T$-system 
(a system of functional relations among transfer matrices) for 
the quantum transfer matrix (QTM) 
\cite{S85,SI87,K87,SAW90,K92}, which 
is independent of the string hypothesis. 
Later on, TBA equations for the $XXZ$-model in the regime 
$|\Delta|< 1$ \cite{KSS98}, 
the $osp(1|2)$ model \cite{ST00}, the $osp(1|2s)$ model \cite{T02} 
were derived by the similar procedures in \cite{JKS98}. 
In addition, the TBA equation in \cite{KSS98} 
was analytically continued \cite{TSK01} to the one for $|\Delta|\ge 1$.  
These TBA equations contain an infinite (or finitely many) number of unknown functions, 
 and thus are not always easily treated.  
It is significant to simplify the TBA equations to tractable 
integral equations which contain only a 
finite number of unknown functions. 

In the context of the QTM method, 
for models related to algebras of rank one or two, 
nonlinear integral equations 
(NLIE) with finite numbers of unknown functions 
 were derived by Kl\"umper and his collaborators 
 \cite{K92,K93,JK97,JKS98-2,S99,FK99}. 
Although their NLIE 
give the same free energy as TBA equations, 
the derivations need trial and error for each model, 
which prevents their extension to higher rank case. 

Another type of NLIE with only one unknown function 
was proposed for the $XXZ$ spin chain by Takahashi \cite{Ta01} recently. 
One can also derive \cite{TSK01} Takahashi's NLIE from 
the $T$-system \cite{KP92,KNS95} of the QTM. 
In view of this fact, we have derived \cite{T02-2} NLIE with a finite 
number of unknown functions from our $T$-system \cite{T99} 
for the $osp(1|2s)$ model for arbitrary rank $s$. 
In this paper, we shall further derive NLIE 
for the $sl(r+1)$ Uimin-Sutherland model \cite{U70,S75} 
with only a {\em finite} number (the number of rank $r$) 
of unknown functions 
from the $T$-system \cite{KNS95}. 
This is the first explicit derivation of this type of NLIE 
for a vertex model associated with $sl(r+1)$ for {\em arbitrary} rank $r$ 

In section 2, we introduce the $sl(r+1)$ Uimin-Sutherland model, 
and define the QTM 
 and the $T$-system \cite{KNS95} 
related to this model. 
In section 3, we derive traditional TBA equations 
from the $T$-system defined in section 2 
without using the string hypothesis. 
In section 4, we derive the NLIE (\ref{nlie3}), (\ref{nlie4}), 
which are our main results.  
The normalized fused QTM $\{{\mathcal T}^{(a)}_{m}(v)\}$ 
($a\in \{1,2,\dots,r\}$; $m\in {\mathbb Z}_{\ge 1}$: 
the fusion degree of the model) 
in the Trotter limit $N \to \infty$ defined in section 2 play the role of 
the unknown functions of these NLIE. 
The traditional TBA equations 
(\ref{TBA-1}), (\ref{TBA-2}), (\ref{TBA-3}) are integral equations 
on the variables $\{Y^{(a)}_{m}(v)\}$, which connects with
 the ones $\{{\mathcal T}^{(a)}_{m}(v)\}$ 
 for our new NLIE (\ref{nlie3}) through the 
 relation (\ref{Y-fun}) (cf. ${\mathcal T}^{(a)}_{m}(v)=
 \lim_{N \to \infty }\widetilde{T}^{(a)}_{m}(v)$). 
In this sense, we may viewed our new NLIE (\ref{nlie3}) 
as TBA equations on 
the variables $\{{\mathcal T}^{(a)}_{m}(v)\}$. 
For the evaluation of the free energy, 
we need only the fundamental one ${\mathcal T}^{(1)}_{1}(v)$. 
Therefore adopting an infinite number of fused QTM 
$\{{\mathcal T}^{(a)}_{m}(v)\}$ as unkown functions 
seems to be superfluous. 
In fact we see that these NLIE (\ref{nlie3}) 
for $m=1$ form a closed set of equations (\ref{nlie4}), 
which contains only a {\em finite} number of unknown functions 
$\{{\mathcal T}^{(a)}_{1}(v)\}_{1\le a \le r}$. 
For $r=1$, 
this set of NLIE (\ref{nlie4}) reduces to 
Takahashi's NLIE \cite{Ta01} for the $XXX$ spin chain.
On the other hand, NLIE (\ref{nlie3}) for $m \ge 2$ 
have never been considered before, and then they are new equations 
 even in the case $r=1$. 
Using our new NLIE (\ref{nlie4}), we calculate the 
high temperature expansion of the free energy in section 5. 
Section 6 are devoted to concluding remarks. 
Many calculations in this paper are parallel with the ones for 
the $osp(1|2s)$ case \cite{T02,T02-2}; 
but we will describe calculations concisely, 
not so much for reader's convenience, 
but for basic importance of the Uimin-Sutherland model. 
\section{$T$-system and QTM method}
We shall introduce the $sl(r+1)$ 
Uimin-Sutherland model \cite{U70,S75}, 
and  define 
the QTM \cite{S85,SI87,K87,SAW90,K92} 
and the $T$-system \cite{KNS95} for this model. 
QTM analyses of the Uimin-Sutherland model 
can be found in \cite{JKS98,FK99}. 
The classical counter part of the $sl(r+1)$ Uimin-Sutherland model
 is a special case of the
  Perk-Schultz model, whose $R$-matrix\cite{PS81} 
is given as 
\begin{eqnarray}
R^{ab}_{\mu \nu}(v)=v\delta_{ab}\delta_{\mu \nu}+
 \delta_{a\nu}\delta_{\mu b}, \label{R-mat}
\end{eqnarray}
where $v\in {\mathbb C}$; 
$a,b,\nu,\mu \in \{1,2,\dots,r+1\} $. 
%
We define the QTM $t^{(1)}_{1}(v)=((t^{(1)}_{1})
^{\{\beta_{1},\dots, \beta_{N} \}}
_{\{\alpha_{1},\dots,\alpha_{N} \}}(v))$ 
as 
\begin{eqnarray}
\hspace{-8pt}
(t^{(1)}_{1})^{\{\beta_{1},\dots, \beta_{N} \}}
_{\{\alpha_{1},\dots,\alpha_{N} \}}(v)=
\sum_{\{\nu_{k}\}}\exp(\frac{\mu_{\nu_{1}}}{T})
\prod_{k=1}^{\frac{N}{2}}
 R^{\nu_{2k-1},\nu_{2k}}_{\alpha_{2k-1},\beta_{2k-1}}(u+iv)
 \widetilde{R}^{\nu_{2k},\nu_{2k+1}}_{\alpha_{2k},\beta_{2k}}(u-iv),
 \label{QTM}
\end{eqnarray}
where $\widetilde{R}^{\mu,\nu}_{\alpha,\beta}(v)=
 R^{\beta,\alpha}_{\mu,\nu}(v)$; 
$N$ is the Trotter number and assumed to be even; 
$u=-\frac{J}{T N}$ ($T$ is a temperature; 
$J$ is a coupling constant, where $J>0$ (resp. $J<0$) 
corresponds to the anti-ferromagnetic (resp. ferromagnetic) regime); 
$\{\mu_{a}\}$ are chemical potentials; 
the Boltzmann constant is set to $1$. 
The free energy per site is expressed 
in terms of  the largest eigenvalue $T^{(1)}_{1}(0)$ of 
the QTM (\ref{QTM}) at $v=0$:
\begin{eqnarray}
f=
-T\lim_{N\to \infty}\log T^{(1)}_{1}(0).
\end{eqnarray} 
The eigenvalue formula $T_{1}^{(1)}(v)$ of the QTM (\ref{QTM}) 
is imbedded into the one $T_{m}^{(a)}(v)$
for a fusion hierarchy of the QTM 
(cf. Bazhanov-Reshetikhin formula in \cite{BR90}): 
\begin{eqnarray}
T_{m}^{(a)}(v)=\sum_{\{d_{j,k}\}} \prod_{j=1}^{a}\prod_{k=1}^{m}
z(d_{j,k};v-\frac{i}{2}(a-m-2j+2k)),
\label{DVF}
\end{eqnarray}
where the summation is taken over 
$d_{j,k}\in 
\{1,2,\dots,r+1\}$ 
such that $d_{j,k} \prec d_{j+1,k}$ and $d_{j,k} \preceq d_{j,k+1}$ 
($ 1 \prec 2 \prec \cdots \prec r+1$). 
The functions $\{z(a;v)\}$ are defined as
\begin{eqnarray}
&& z(a;v)=\psi_{a}(v)
\frac{Q_{a-1}(v-\frac{i}{2}(a+1))Q_{a}(v-\frac{i}{2}(a-2))}
{Q_{a-1}(v-\frac{i}{2}(a-1))Q_{a}(v-\frac{i}{2}a)} 
\nonumber \\ 
&& \hspace{150pt} 
\mbox{for} \quad a \in \{1,2,\dots,r+1\}, 
\end{eqnarray}
where $Q_{a}(v)=\prod_{k=1}^{M_{a}}(v-v_{k}^{(a)})$; 
$M_{a}\in {\mathbb Z}_{\ge 0}$; $Q_{0}(v)=Q_{r+1}(v)=1$.  
The vacuum parts are given as follows
\begin{eqnarray}
\psi_{a}(v)=
 e^{\frac{\mu_{a}}{T}}
 \phi_{+}(v+i \delta_{a,r+1})\phi_{-}(v-i\delta_{a,1})
 \quad 
 \mbox{for} \quad a \in \{1,2,\dots,r+1\},
 \label{vac-QTM}
\end{eqnarray}
where $\phi_{\pm}(v)=(v\pm iu)^{\frac{N}{2}}$. 
$\{v^{(a)}_{k}\}$ is a solution of the Bethe ansatz equation 
(BAE)
\begin{eqnarray}
&& \frac{\psi_{a}(v^{(a)}_{k}+\frac{i}{2}a)}
     {\psi_{a+1}(v^{(a)}_{k}+\frac{i}{2}a)}=
-
\frac{Q_{a-1}(v^{(a)}_{k}+\frac{i}{2})Q_{a}(v^{(a)}_{k}-i)
      Q_{a+1}(v^{(a)}_{k}+\frac{i}{2})}
      {Q_{a-1}(v^{(a)}_{k}-\frac{i}{2})Q_{a}(v^{(a)}_{k}+i)
      Q_{a+1}(v^{(a)}_{k}-\frac{i}{2})}
     \label{BAE} \\
&& \hspace{40pt} \mbox{for} \quad k\in \{1,2, \dots, M_{a}\} \quad 
\mbox{and} \quad a\in \{1,2, \dots, r\}.
\nonumber
\end{eqnarray}
We can rewrite (\ref{BAE}) as the Reshetikhin and Wiegmann's 
BAE \cite{RW87} 
in terms of the representation theoretical data.
\begin{eqnarray}
&& -\prod_{j=1}^{N}
\left(
\frac{v^{(a)}_{k}-\frac{i}{2}b^{(a)}_{j}-w^{(a)}_{j}}
 {v^{(a)}_{k}+\frac{i}{2}b^{(a)}_{j}-w^{(a)}_{j}}
\right)
 =\zeta_{a}
 \prod_{b=1}^{r}\frac{Q_{b}(v^{(a)}_{k}-\frac{i}{2}(\alpha_{a}|\alpha_{b}))}
 {Q_{b}(v^{(a)}_{k}+\frac{i}{2}(\alpha_{a}|\alpha_{b}))}\\
 &&\hspace{40pt} \mbox{for} \quad a \in \{1,2,\dots,r\}, \nonumber 
\end{eqnarray}
where 
$(\alpha_{a}|\alpha_{b})=2\delta_{a,b}-\delta_{a,b+1}-\delta_{a,b-1}$ 
is the Cartan matrix of $sl(r+1)$; 
$\{ b^{(a)}_{j} \}$ are the Dynkin labels which characterize 
the quantum space: 
$[b^{(1)}_{2j},b^{(2)}_{2j},\dots,b^{(r)}_{2j}]=[0,0,\dots,1]$, 
$[b^{(1)}_{2j-1},b^{(2)}_{2j-1},\dots,b^{(r)}_{2j-1}]=[1,0,\dots,0]$; 
$\{w^{(a)}_{j}\}$ are inhomogeneity parameters: 
$w^{(a)}_{2j-1}=\delta_{a,1}iu$, 
$w^{(a)}_{2j}=-\delta_{a,r}i(u+\frac{1}{2}g)$;
$g=r+1$ (the dual Coxeter number); 
$\zeta_{a}=e^{-\frac{\mu_{a}-\mu_{a+1}}{T}}$. 
Note that 
the Dynkin labels for the even site and the odd site are 
the ones for conjugate representations each other. 
The non-negative integers $\{M_{a}\}$ should satisfy the condition: 
$\sum_{j=1}^{N}b^{(a)}_{j}-\sum_{b=1}^{r}M_{b}(\alpha_{b}|\alpha_{a})
=\frac{N}{2}(\delta_{a,1}+\delta_{a,r})+M_{a-1}-2M_{a}+M_{a+1}
\in {\mathbb Z}_{\ge 0}$, where $M_{0}=M_{r+1}=0$. 
For the row-to-row transfer matrix, the above parameters 
take the values: 
$[b^{(1)}_{j},b^{(2)}_{j},\dots,b^{(r)}_{j}]=[1,0,\dots,0]$; 
$w^{(a)}_{j}=0$; $\zeta_{a}=1$. 

For $a \in \{1,2,\dots,r\}$ and $m \in {\mathbb Z}_{\ge 1}$, we 
 shall normalize (\ref{DVF}) as 
 $ \widetilde{T}^{(a)}_{m}(v)=
 T^{(a)}_{m}(v)/\widetilde{{\mathcal N}}^{(a)}_{m}(v)$, 
 where 
\begin{eqnarray}
\hspace{-30pt} && \widetilde{{\mathcal N}}^{(a)}_{m}(v)=
  \frac{\phi_{-}(v-\frac{a+m}{2}i)\phi_{+}(v+\frac{a+m}{2}i)}{
  \phi_{-}(v-\frac{a-m}{2}i)\phi_{+}(v+\frac{a-m}{2}i)}
  \nonumber \\ 
\hspace{-30pt}  && \hspace{20pt} \times
  \prod_{j=1}^{a}\prod_{k=1}^{m}
  \phi_{-}(v-\frac{a-m-2j+2k}{2}i)\phi_{+}(v-\frac{a-m-2j+2k}{2}i).
  \label{normal}
\end{eqnarray}
Note that the poles of $\widetilde{T}^{(a)}_{m}(v)$ 
from the functions $\{Q_{b}(v) \}$ (dress part) 
are spurious under the BAE (\ref{BAE}).  
One can show that 
$\widetilde{T}^{(a)}_{m}(v)$ satisfies 
the following functional relation.  
\begin{eqnarray}
&& \hspace{-20pt}
\widetilde{T}^{(a)}_{m}(v+\frac{i}{2})
\widetilde{T}^{(a)}_{m}(v-\frac{i}{2})
=\widetilde{T}^{(a)}_{m+1}(v)\widetilde{T}^{(a)}_{m-1}(v)
+\widetilde{T}^{(a-1)}_{m}(v)\widetilde{T}^{(a+1)}_{m}(v),
\label{T-system}
\\ && \hspace{70pt} 
{\rm for} \quad a \in \{1,2,\dots,r\}
\quad {\rm and} \quad m \in {\mathbb Z}_{\ge 1}, 
\nonumber
\end{eqnarray}
where 
\begin{eqnarray}
&& \widetilde{T}^{(a)}_{0}(v)=1
\quad {\rm for} \quad a \in {\mathbb Z}_{\ge 1},\nonumber \\
&& \widetilde{T}^{(0)}_{m}(v)=
 \frac{\phi_{-}(v+\frac{m}{2}i)\phi_{+}(v-\frac{m}{2}i)}
  {\phi_{-}(v-\frac{m}{2}i)\phi_{+}(v+\frac{m}{2}i)}
\quad {\rm for} \quad m \in {\mathbb Z}_{\ge 1}, \\
&& \widetilde{T}^{(r+1)}_{m}(v)=
e^{\frac{m(\mu_{1}+\mu_{2}+\cdots +\mu_{r+1})}{T}} 
\quad {\rm for} \quad m \in {\mathbb Z}_{\ge 1} \nonumber. 
\end{eqnarray}
Except for the vacuum part
 (the part made from the function (\ref{vac-QTM})), 
this equation has the 
same form as the $sl(r+1)$ 
$T$-system in \cite{KNS95}. 
\section{Traditional TBA equations}
In this section, we will transform the 
$T$-system (\ref{T-system}) into 
the traditional TBA equations by the standard procedure
\cite{KP92,JKS98}. 
Similar argument in relation to Stokes multipliers 
can be found in section 5 in \cite{S00}. 

For $m \in {\mathbb Z}_{\ge 1}$ and $a \in \{1,2,\dots,r\}$, 
we define functions:
\begin{eqnarray}
Y^{(a)}_{m}(v)=
 \frac{\widetilde{T}^{(a)}_{m+1}(v)\widetilde{T}^{(a)}_{m-1}(v)}
 {\widetilde{T}^{(a-1)}_{m}(v)\widetilde{T}^{(a+1)}_{m}(v)}
 =\frac{T^{(a)}_{m+1}(v)T^{(a)}_{m-1}(v)}
 {T^{(a-1)}_{m}(v)T^{(a+1)}_{m}(v)}. \label{Y-fun}
\end{eqnarray}
By using the $T$-system (\ref{T-system}), we can show that 
(\ref{Y-fun}) satisfy the following $Y$-system:
\begin{eqnarray}
 && Y^{(a)}_{m}(v+\frac{i}{2})Y^{(a)}_{m}(v-\frac{i}{2})=
 \frac{(1+Y^{(a)}_{m+1}(v))(1+Y^{(a)}_{m-1}(v))}
 {\prod_{b=1}^{r}(1+(Y_{m}^{(b)}(v))^{-1})^{I_{ab}}}
 \label{Y-sys} \\
&& \hspace{40pt} \mbox{for} \quad 
a \in \{1,2,\dots,r\} \quad \mbox{and} \quad 
m \in {\mathbb Z}_{\ge 1}, \nonumber 
\end{eqnarray}
where $I_{ab}=\delta_{a,b+1}+\delta_{a,b-1}$, 
$Y^{(a)}_{0}(v)=0$. 

\begin{figure}
\includegraphics[width=0.95\textwidth]{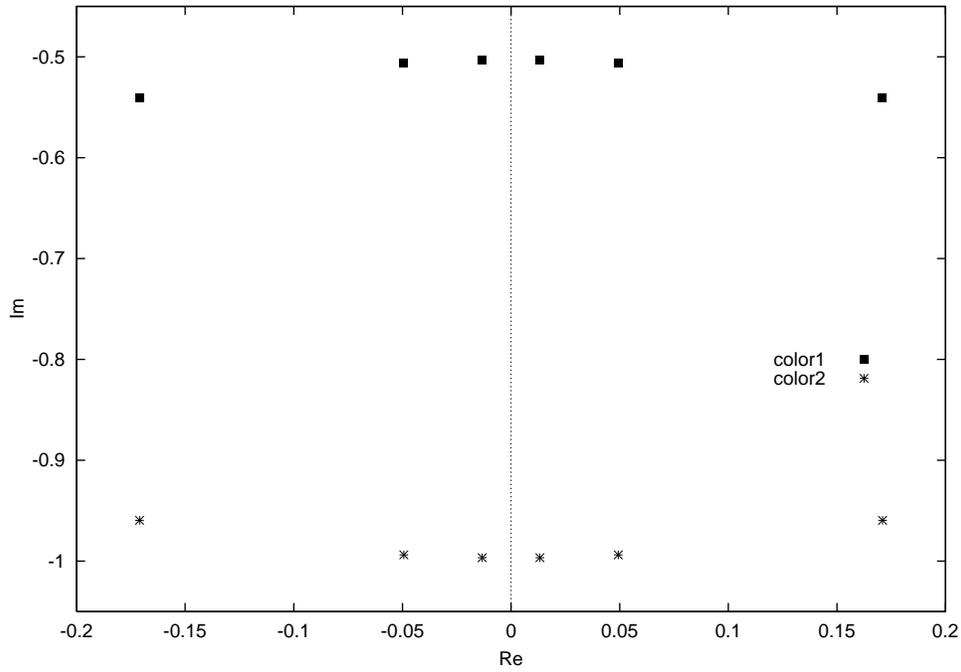}
\caption{Location of the roots of the BAE 
for $sl(3)$ case ($N=12$,$u=-0.05$, 
$\mu_{1}=\mu_{2}=\mu_{3}=0$), which gives 
the largest eigenvalue of the QTM $t^{(1)}_{1}(v)$ 
at $v=0$. 
Both color 1 roots $\{v^{(1)}_{k}\}$ 
and color 2 roots $\{v^{(2)}_{k}\}$ 
form six one-strings.}
\label{roots}
\end{figure}

From a numerical analysis for finite $N,u,r$, 
we expect that  
a one-string solution (for every color) in the sector 
$\frac{N}{2}=M_{1}=M_{2}=\cdots =M_{r}$ 
of the BAE (\ref{BAE}) provides 
the largest eigenvalue of the QTM (\ref{QTM}) at $v=0$  
at least for the case: $\mu_{1}=\mu_{2}=\dots =\mu_{r+1}=0$ 
(cf. Figure \ref{roots}). 
Hereafter we will consider only this one-string solution. 
\begin{figure}
\includegraphics[width=0.95\textwidth]{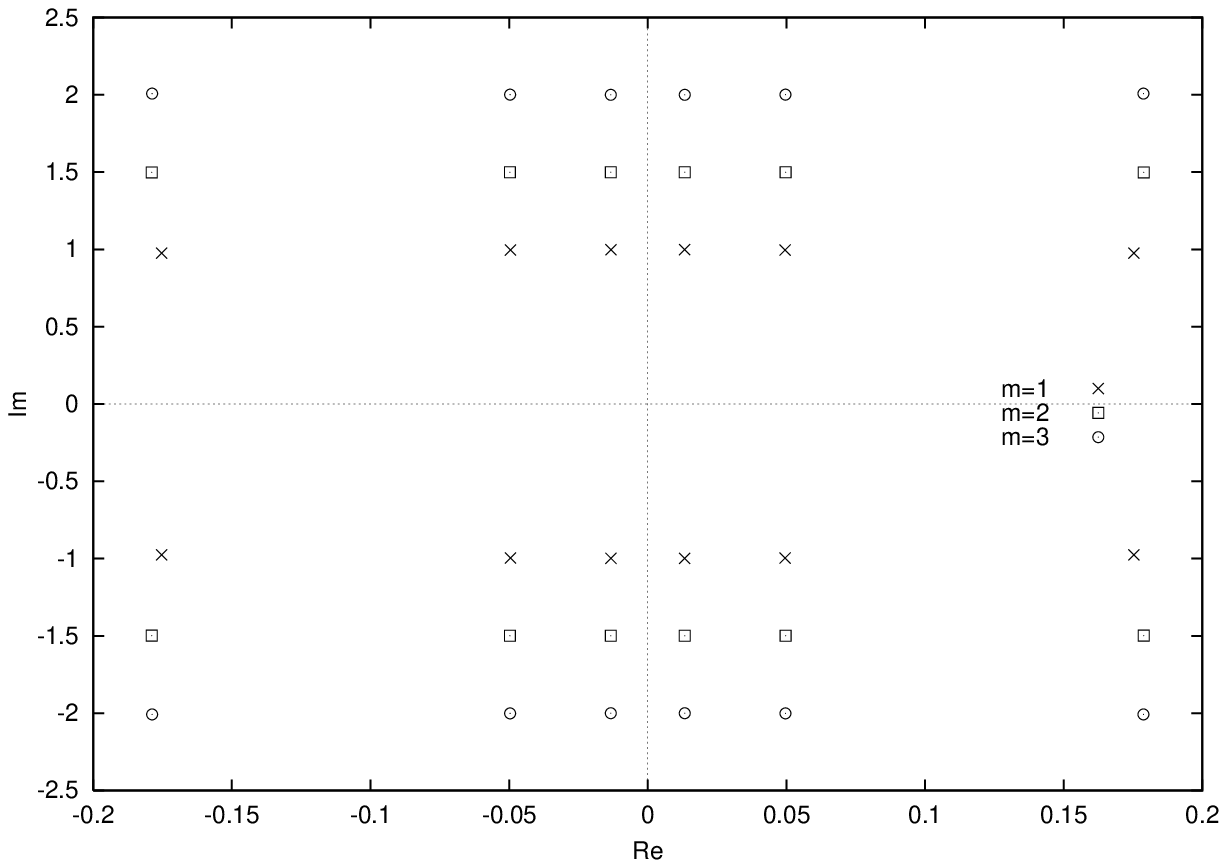}
\caption{Location of zeros of $\widetilde{T}^{(1)}_{m}(v)$ 
for $sl(3)$ case ($m=1,2,3$, $N=12$,$u=-0.05$, 
$\mu_{1}=\mu_{2}=\mu_{3}=0$).
The zeros recede from the physical strip 
${\rm Im}v \in [-\frac{1}{2},\frac{1}{2}]$ as $m$ increases.}
\label{zeros1}
\end{figure}
\begin{figure}
\includegraphics[width=0.95\textwidth]{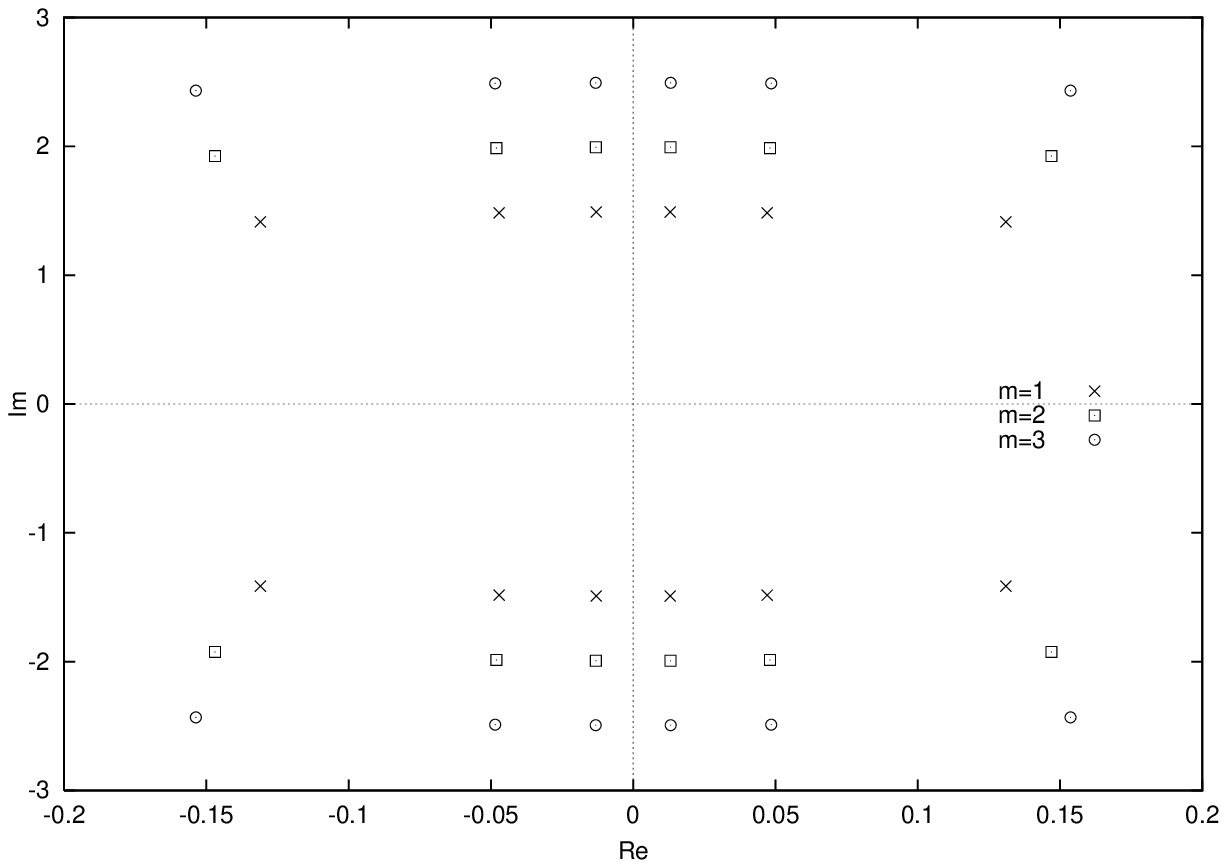}
\caption{Location of zeros of $\widetilde{T}^{(2)}_{m}(v)$ 
for $sl(3)$ case ($m=1,2,3$, $N=12$,$u=-0.05$,
 $\mu_{1}=\mu_{2}=\mu_{3}=0$).
The zeros recede from the physical strip 
${\rm Im}v \in [-\frac{1}{2},\frac{1}{2}]$ as $m$ increases.}
\label{zeros2}
\end{figure}
The following conjecture will be valid for 
this one-string solution (cf. Figures \ref{zeros1}, \ref{zeros2}). 
\begin{conjecture}\label{conj}
For small $u$ ($|u|\ll 1$), $a \in \{1,2,\dots,r\}$ and 
$m \in {\mathbb Z}_{\ge 1}$,
every zero $\{\tilde{z}^{(a)}_{m} \}$ 
of $\widetilde{T}^{(a)}_{m}(v)$ is located near the lines 
${\mathrm Im} v= \pm \frac{m+a}{2}$ 
at least for the case: $\mu_{1}=\mu_{2}=\dots =\mu_{r+1}=0$. 
\end{conjecture}
Based on this conjecture, 
we will establish the ANZC property 
(Analytic, NonZero and Constant asymptotics in 
the limit $|v| \to \infty$) 
in some domain for the functions (\ref{Y-fun}). 
We find that $Y^{(a)}_{m}(v)$ has a constant asymptotics:
\begin{eqnarray}
\lim_{|v| \to \infty}Y^{(a)}_{m}(v)=
\frac{Q^{(a)}_{m-1}Q^{(a)}_{m+1}}{Q^{(a-1)}_{m}Q^{(a+1)}_{m}},
\label{limit}
\end{eqnarray}
where $\{Q^{(a)}_{m}\}$ is given by (\ref{DVF}) 
with a formal setting $Q_{a}(v)=\phi_{\pm}(v)=1$. 
$\{Q^{(a)}_{m}\}$ is also characterized as 
a solution of the $Q$-system \cite{KR90}: 
\begin{eqnarray}
&& \hspace{-40pt}
(Q_{m}^{(a)})^{2}=Q_{m-1}^{(a)}Q_{m+1}^{(a)}
 +Q_{m}^{(a-1)}Q_{m}^{(a+1)} 
 \ {\rm for} \ a \in \{1,2,\dots, r\} 
 \ \mbox{and} \ m \in {\mathbb Z}_{\ge 1}, 
 \label{Q-sys}
\end{eqnarray}
where  
$Q^{(a)}_{0}=Q^{(0)}_{m}=1$; 
$Q^{(r+1)}_{m}=e^{\frac{m(\mu_{1}+\mu_{2}+\cdots +\mu_{r+1})}{T}}$. 
Note that $Q^{(a)}_{m}$ coincides with the character  
of $m$-th symmetric $a$-th 
anti-symmetric tensor representation $V(m\Lambda_{a})$ of $gl(r+1)$ 
if we set $\frac{\mu_{a}}{T} \to \epsilon_{a}$, 
where $\{ \epsilon_{a}\}$ are orthonormal basis of 
the dual space of the Cartan subalgebra. 
In particular, for $\mu_{1}=\mu_{2}=\cdots =\mu_{r+1}=0$ case, 
$Q^{(a)}_{m}$ is given by 
\begin{eqnarray}
Q^{(a)}_{m}=\left(\frac{(m+g)!m!}{(m+a)!(m+g-a)!}\right)^m
 \prod_{k=1}^{m}\left\{ \frac{(k+a)(k+g-a)}{k(k+g)} \right\}^{k},
 \label{q-sys-sol}
\end{eqnarray}
and (\ref{limit})
has the form
\begin{eqnarray}
\lim_{|v| \to \infty}Y^{(a)}_{m}(v)=\frac{m(g+m)}{a(g-a)}.
\label{limit2}
\end{eqnarray}
(\ref{q-sys-sol}) is the dimension of $V(m\Lambda_{a})$. 
Based on the conjecture \ref{conj} and (\ref{limit}), 
one can see that the functions $1+Y_{m}^{(a)}(v)$, 
$1+(Y_{m}^{(a)}(v))^{-1}$ 
 in the domain ${\rm Im}v \in [-\delta,\delta]$ 
($0<\delta \ll 1$) and  
$Y^{(a)}_{m}(v)$ for $(a,m)\ne (1,1)$ in the domain  
${\rm Im}v \in [-\frac{1}{2},\frac{1}{2}]$ (physical strip) 
have the ANZC property at least for 
the case $\mu_{1}=\cdots =\mu_{r+1}=0$; while  
$Y^{(1)}_{1}(v)$ has zeros of order $N/2$ 
at $\pm i(\frac{1}{2}+u)$ if $u<0$ ($J>0$), 
poles of order $N/2$ 
at $\pm i(\frac{1}{2}-u)$ if $u>0$ ($J<0$) in the physical strip. 
Thus one should modify $Y^{(a)}_{m}(v)$ as 
\begin{eqnarray}
\hspace{-10pt}
\widetilde{Y}^{(a)}_{m}(v)=Y^{(a)}_{m}(v)\left\{
\tanh\frac{\pi}{2}(v+i(\frac{1}{2} \pm u))\tanh\frac{\pi}{2}
(v-i(\frac{1}{2} \pm u))\right\}^{\mp \frac{N\delta_{a,1}\delta_{m,1}}{2}}
\hspace{-20pt} ,
\end{eqnarray}
where the sign $\pm $ is equal to that of $-u$. 
By using the relation
\begin{eqnarray}
\tanh\frac{\pi}{4}(v+i)\tanh\frac{\pi}{4}(v-i)=1,
\end{eqnarray}
we shall modify the lhs of the $Y$-system (\ref{Y-sys}) as
\begin{eqnarray} 
&& \widetilde{Y}^{(a)}_{m}(v-\frac{i}{2})\widetilde{Y}^{(a)}_{m}(v+\frac{i}{2})
 =\frac{(1+Y^{(a)}_{m+1}(v))(1+Y^{(a)}_{m-1}(v))}
 {\prod_{b=1}^{r}(1+(Y_{m}^{(b)}(v))^{-1})^{I_{ab}}}, 
 \label{modi-Y-sys} \\
&& \hspace{120pt}{\rm for} \quad m \in {\mathbb Z}_{\ge 1} 
\quad {\rm and} \quad a\in \{1,2,\dots,r\}.\nonumber 
\end{eqnarray}
This modified $Y$-system has the ANZC property.  
Then one can transform (\ref{modi-Y-sys}) into 
nonlinear integral equations 
by a standard procedure. 
\begin{eqnarray}
\log Y_{m}^{(a)}(v)&=&
\pm \frac{N\delta_{a1}\delta_{m1}}{2}
\log \left\{
\tanh\frac{\pi}{2}(v+i(\frac{1}{2} \pm u))\tanh\frac{\pi}{2}
(v-i(\frac{1}{2} \pm u))
\right\} \nonumber \\
&& +K*\log\left\{
\frac{(1+Y_{m-1}^{(a)})(1+Y_{m+1}^{(a)})}
{\prod_{b=1}^{r}(1+(Y_{m}^{(b)})^{-1})^{I_{ab}}}\right\}(v) 
\label{nonlinear}
\\
&& \hspace{30pt}
 \mbox{for} \quad a \in \{1,2,\dots,r\} \quad 
 \mbox{and} \quad 
 m \in {\mathbb Z}_{\ge 1}, 
 \nonumber
\end{eqnarray}
where $Y^{(a)}_{0}(v)=0$ and $*$ is the convolution.
After the Trotter limit $N \to \infty$ with $u=-\frac{J}{NT}$, 
we obtain a TBA equation 
\begin{eqnarray}
&& \hspace{-30pt} \log Y_{m}^{(a)}(v)=
-\frac{\pi J \delta_{a1}\delta_{m1}}{T \cosh\pi v}
+K*\log\left\{
\frac{(1+Y_{m-1}^{(a)})(1+Y_{m+1}^{(a)})}
{\prod_{b=1}^{r}(1+(Y_{m}^{(b)})^{-1})^{I_{ab}}}\right\}(v)
\label{TBA-1}
\\
&& \hspace{40pt}
 \mbox{for} \quad a \in \{1,2,\dots,r\} \quad 
 \mbox{and} \quad 
 m \in {\mathbb Z}_{\ge 1}, 
 \nonumber
\end{eqnarray}
where 
 $Y_{0}^{(a)}(v):=0$ and the kernel is 
\begin{eqnarray}
K(v)=\frac{1}{2 \cosh \pi v}. 
\end{eqnarray}
One can also rewrite (\ref{TBA-1}) as 
\begin{eqnarray}
&& -\frac{\pi J \delta_{a1}\delta_{m1}}{T \cosh\pi v}=
\sum_{n=1}^{\infty} K^{m,n}*\log (1+Y_{n}^{(a)})(v) 
 \label{TBA-2} \\
&& \hspace{80pt} -\sum_{b=1}^{r} \sum_{n=1}^{\infty} 
   J^{m,n}_{a,b}*\log (1+(Y_{n}^{(b)})^{-1})(v),
\nonumber
\end{eqnarray}
where 
\begin{eqnarray}
&& K^{m,n}(v)=\delta_{m,n}\delta(v)
  -(\delta_{m,n-1}+\delta_{m,n+1})K(v), \nonumber \\ 
&& J^{n,k}_{a,b}(v)=(\delta_{a,b}\delta(v)
  -(\delta_{a,b-1}+\delta_{a,b+1})K(v))\delta_{n,k}.
\end{eqnarray}
We find that 
the TBA equation (\ref{TBA-2}) coincides with the 
one from the string hypothesis in \cite{BR90} after 
suitable modification. 
By using the following relations
\begin{eqnarray}
&& A^{a,d}(v)=\sum_{l=1}^{\min(a,d)}G_{|a-d|+2l-1}(v) ,\nonumber \\
&& G_{a}(v)=\frac{1}{g}
\frac{\sin\frac{(g-a)\pi}{g}}
{\cos \frac{(g-a)\pi}{g} + \cosh\frac{2\pi v}{g}}, \nonumber \\
&& \widehat{A}^{a,d}(k)=\int_{-\infty}^{\infty}{\mathrm d}v
 A^{a,d}(v)e^{-ikv} 
 =\frac{\sinh \left(\frac{\min (a,d)}{2} k\right) 
    \sinh \left(\frac{g-\min (a,d)}{2} k\right)}
    {\sinh \left(\frac{k}{2}\right) 
    \sinh \left(\frac{g}{2} k\right)}, \nonumber  \\
&& \sum_{l=1}^{r}\widehat{A}^{m,l}(k)\widehat{K}^{l,n}(k)
= \frac{\delta_{m,n}}{2\cosh(\frac{k}{2})},
  \label{relations} \\ 
&& \widehat{K}^{m,n}(k)=
\delta_{m,n}-\frac{\delta_{m,n-1}+\delta_{m,n+1}}{2\cosh\frac{k}{2}},\nonumber
\end{eqnarray}
 we can also rewrite the TBA equation (\ref{TBA-1}) as
\begin{eqnarray}
\log Y_{m}^{(a)}(v)&=& -\frac{2\pi J \delta_{m,1}}{T}G_{a}(v) \nonumber \\ 
&& +\sum_{b=1}^{r}A^{a,b}*\log\left\{
\frac{(1+Y_{m-1}^{(b)})(1+Y_{m+1}^{(b)})}
{\prod_{d=1}^{r}(1+Y_{m}^{(d)})^{I_{bd}}}\right\}(v)
\label{TBA-3}
\\
&& \hspace{30pt}
 \mbox{for} \quad a \in \{1,2,\dots,r\} \quad 
 \mbox{and} \quad 
 m \in {\mathbb Z}_{\ge 1}, 
 \nonumber
\end{eqnarray}
where $Y^{(a)}_{0}(v)=0$. 
This TBA equation (\ref{TBA-3}) does not contain $(Y^{(a)}_{m})^{-1}$, 
which contrasts with (\ref{TBA-1}),(\ref{TBA-2}). 
We can derive the following relation 
from (\ref{T-system}), (\ref{Y-fun}) and (\ref{relations}).
\begin{eqnarray}
\log \widetilde{T}^{(1)}_{1}(v)&=&
  \sum_{b=1}^{r}G_{b}*\log(1+Y^{(b)}_{1})(v) 
 \nonumber \\
&& -2N \int_{0}^{\infty}{\mathrm d}k 
 \frac{e^{-\frac{1}{2}k}\sinh(ku)\cos(kv)\sinh(\frac{g-1}{2}k)}{
 k \sinh(\frac{g}{2}k)}.
\end{eqnarray}
After the Trotter limit $N \to \infty$ with $u=-\frac{J}{NT}$, 
we have 
the free energy per site. 
\begin{eqnarray}
f &=&-T\lim_{N \to \infty}\log T^{(1)}_{1}(0)=
J-T\lim_{N \to \infty}\log \widetilde{T}^{(1)}_{1}(0) 
\label{free-finite}\\
&=&
J\left\{
\frac{2}{g}\left(
\gamma +\psi(\frac{1}{g})
\right)
+1 
\right\} -T
\sum_{a=1}^{r}
\int_{-\infty}^
{\infty}{\mathrm d}v G_{a}(v)\log(1+Y^{(a)}_{1}(v)), 
\nonumber 
\end{eqnarray}
where $\psi(z)$ is the digamma function
\begin{eqnarray}
\psi(z)&=&\frac{d}{dz}\log \Gamma (z), 
\qquad 
\gamma=-\psi(1).
\end{eqnarray}
\section{New nonlinear integral equations}
It was pointed out \cite{TSK01} that Takahashi's NLIE 
for the $XXZ$-model \cite{Ta01} can be rederived from 
the $T$-system of the QTM. 
In this section, we shall derive our new 
NLIE (\ref{nlie3}) and (\ref{nlie4}) from 
the $T$-system (\ref{T-system}). 

$\widetilde{T}^{(a)}_{m}(v)$ has poles only at 
$\pm \tilde{\beta}^{(a)}_{m}$: 
$\tilde{\beta}^{(a)}_{m}=\frac{m+a}{2}i +iu$. 
These poles are of order $N/2$ at most. 
In addition,  
$\lim_{|v|\to \infty}\widetilde{T}^{(a)}_{m}(v)=Q^{(a)}_{m}$ 
is a finite number (cf. (\ref{q-sys-sol})). 
Thus we must put 
\begin{eqnarray}
\widetilde{T}^{(a)}_{m}(v)=Q^{(a)}_{m} +
\sum_{j=1}^{\frac{N}{2}}
\left\{
  \frac{b^{(a)}_{m,j}}{(v-\tilde{\beta}^{(a)}_{m})^{j}}
 +\frac{\overline{b}^{(a)}_{m,j}}{(v+\tilde{\beta}^{(a)}_{m})^{j}}
\right\},
\label{expan}
\end{eqnarray}
where the coefficients $b^{(a)}_{m,j},
\overline{b}^{(a)}_{m,j} \in {\mathbb C}$ 
are given as follows.
\begin{eqnarray}
&& b^{(a)}_{m,j}= \oint_{C^{(a)}_{m}} \frac{{\mathrm d} v}{2\pi i}
 \widetilde{T}^{(a)}_{m}(v)(v-\tilde{\beta}^{(a)}_{m})^{j-1},\nonumber \\
&& \overline{b}^{(a)}_{m,j}=
 \oint_{\overline{C}^{(a)}_{m}} \frac{{\mathrm d} v}{2\pi i}
 \widetilde{T}^{(a)}_{m}(v)(v+\tilde{\beta}^{(a)}_{m})^{j-1},
 \label{coeff}
\end{eqnarray}
where the contour $C^{(a)}_{m}$ is a counterclockwise 
closed loop around 
$\tilde{\beta}^{(a)}_{m}$ 
 which does not surround $-\tilde{\beta}^{(a)}_{m}$; 
the contour $\overline{C}^{(a)}_{m}$ is
a counterclockwise 
closed loop around $-\tilde{\beta}^{(a)}_{m}$ 
which does not surround $\tilde{\beta}^{(a)}_{m}$.
Using the $T$-system (\ref{T-system}), one can modify (\ref{coeff}) as 
\begin{eqnarray}
&& b^{(a)}_{m,j}= \oint_{C^{(a)}_{m}} \frac{{\mathrm d} v}{2\pi i}
 \bigg\{
 \frac{\widetilde{T}^{(a)}_{m-1}(v-\frac{i}{2})
       \widetilde{T}^{(a)}_{m+1}(v-\frac{i}{2})}
      {\widetilde{T}^{(a)}_{m}(v-i)} \nonumber \\
&& \hspace{80pt} +
 \frac{\widetilde{T}^{(a-1)}_{m}(v-\frac{i}{2})
       \widetilde{T}^{(a+1)}_{m}(v-\frac{i}{2})}
      {\widetilde{T}^{(a)}_{m}(v-i)}
 \bigg\}
 (v-\tilde{\beta}^{(a)}_{m})^{j-1},\nonumber \\
&& \overline{b}^{(a)}_{m,j}=
 \oint_{\overline{C}^{(a)}_{m}} \frac{{\mathrm d} v}{2\pi i}
\bigg\{
 \frac{\widetilde{T}^{(a)}_{m-1}(v+\frac{i}{2})
       \widetilde{T}^{(a)}_{m+1}(v+\frac{i}{2})}
      {\widetilde{T}^{(a)}_{m}(v+i)} \nonumber \\
&& \hspace{80pt} +
 \frac{\widetilde{T}^{(a-1)}_{m}(v+\frac{i}{2})
       \widetilde{T}^{(a+1)}_{m}(v+\frac{i}{2})}
      {\widetilde{T}^{(a)}_{m}(v+i)}
 \bigg\}
 (v+\tilde{\beta}^{(a)}_{m})^{j-1}. 
 \label{coeff2}
\end{eqnarray}
%
Substituting (\ref{coeff2}) into (\ref{expan}), we obtain
\begin{eqnarray}
&& \hspace{-20pt}
\widetilde{T}^{(a)}_{m}(v)=Q^{(a)}_{m} \nonumber \\ 
&& +
\oint_{C^{(a)}_{m}} \frac{{\mathrm d} y}{2\pi i} 
\frac{1-\left(\frac{y}{v-\tilde{\beta}^{(a)}_{m}}\right)^{\frac{N}{2}}}
 {v-y-\tilde{\beta}^{(a)}_{m}}
 \bigg\{ 
 \frac{\widetilde{T}^{(a)}_{m-1}(y+\tilde{\beta}^{(a)}_{m}-\frac{i}{2}) 
 \widetilde{T}^{(a)}_{m+1}(y+\tilde{\beta}^{(a)}_{m}-\frac{i}{2})}
 {\widetilde{T}^{(a)}_{m}(y+\tilde{\beta}^{(a)}_{m}-i)} \nonumber \\
 && \hspace{90pt} +
 \frac{\widetilde{T}^{(a-1)}_{m}(y+\tilde{\beta}^{(a)}_{m}-\frac{i}{2}) 
 \widetilde{T}^{(a+1)}_{m}(y+\tilde{\beta}^{(a)}_{m}-\frac{i}{2})}
 {\widetilde{T}^{(a)}_{m}(y+\tilde{\beta}^{(a)}_{m}-i)}
 \bigg\} \nonumber \\
&& +
\oint_{\overline{C}^{(a)}_{m}} \frac{{\mathrm d} y}{2\pi i} 
 \frac{1-\left(\frac{y}{v+\tilde{\beta}^{(a)}_{m}}\right)^{\frac{N}{2}}}
 {v-y+\tilde{\beta}^{(a)}_{m}}
 \bigg\{ 
 \frac{\widetilde{T}^{(a)}_{m-1}(y-\tilde{\beta}^{(a)}_{m}+\frac{i}{2}) 
 \widetilde{T}^{(a)}_{m+1}(y-\tilde{\beta}^{(a)}_{m}+\frac{i}{2})}
 {\widetilde{T}^{(a)}_{m}(y-\tilde{\beta}^{(a)}_{m}+i)} \nonumber \\
 && \hspace{90pt} +
 \frac{\widetilde{T}^{(a-1)}_{m}(y-\tilde{\beta}^{(a)}_{m}+\frac{i}{2}) 
 \widetilde{T}^{(a+1)}_{m}(y-\tilde{\beta}^{(a)}_{m}+\frac{i}{2})}
 {\widetilde{T}^{(a)}_{m}(y-\tilde{\beta}^{(a)}_{m}+i)}
 \bigg\}, \nonumber \\ 
 && \hspace{100pt} 
 {\rm for} \quad a \in \{1,2,\dots,r\}
 \quad {\rm and} \quad m \in {\mathbb Z}_{\ge 1},
 \label{nlie1}
\end{eqnarray}
where the contour $C^{(a)}_{m}$ (resp. $\overline{C}^{(a)}_{m}$) 
is a counterclockwise 
closed loop around $0$ 
which does not surround 
$-2\tilde{\beta}^{(a)}_{m}$
(resp. $ 2 \tilde{\beta}^{(a)}_{m}$).
%
The first term and the second term in the first bracket $\{ \cdots \}$ in 
(\ref{nlie1}) have a common singularity at $0$; while for $m=1$, 
this singularity from the first term disappears since 
$\widetilde{T}^{(a)}_{0}(y)=1$. Thus for $m=1$, 
the contribution to the contour integral from 
the first term in the first bracket $\{ \cdots \}$ in 
(\ref{nlie1}) vanishes if 
 the contour $C^{(a)}_{1}$ does not surround the 
singularity
 at $\tilde{z}^{(a)}_{1}-\tilde{\beta}^{(a)}_{1}+i$ 
(cf. Conjecture \ref{conj}). 
This is also the case with the second bracket $\{ \cdots \}$ in 
(\ref{nlie1}).
Therefore for $m=1$, (\ref{nlie1}) reduces to 
\begin{eqnarray}
&& \hspace{-20pt}
\widetilde{T}^{(a)}_{1}(v)=Q^{(a)}_{1} \nonumber \\ 
&& +
\oint_{C^{(a)}_{1}} \frac{{\mathrm d} y}{2\pi i} 
\frac{1-\left(\frac{y}{v-\tilde{\beta}^{(a)}_{1}}\right)^{\frac{N}{2}}}
 {v-y-\tilde{\beta}^{(a)}_{1}}
 \frac{\widetilde{T}^{(a-1)}_{1}(y+\tilde{\beta}^{(a)}_{1}-\frac{i}{2}) 
 \widetilde{T}^{(a+1)}_{1}(y+\tilde{\beta}^{(a)}_{1}-\frac{i}{2})}
 {\widetilde{T}^{(a)}_{1}(y+\tilde{\beta}^{(a)}_{1}-i)}
 \nonumber \\
&& +
\oint_{\overline{C}^{(a)}_{1}} \frac{{\mathrm d} y}{2\pi i} 
 \frac{1-\left(\frac{y}{v+\tilde{\beta}^{(a)}_{1}}\right)^{\frac{N}{2}}}
 {v-y+\tilde{\beta}^{(a)}_{1}}
 \frac{\widetilde{T}^{(a-1)}_{1}(y-\tilde{\beta}^{(a)}_{1}+\frac{i}{2}) 
 \widetilde{T}^{(a+1)}_{1}(y-\tilde{\beta}^{(a)}_{1}+\frac{i}{2})}
 {\widetilde{T}^{(a)}_{1}(y-\tilde{\beta}^{(a)}_{1}+i)}
 \nonumber \\ 
 && \hspace{180pt} 
 {\rm for} \quad a \in \{1,2,\dots,r\}, \label{nlie2} 
\end{eqnarray}
where the contour $C^{(a)}_{1}$ (resp. $\overline{C}^{(a)}_{1}$) 
is a counterclockwise 
closed loop around $0$ which does not surround 
$\tilde{z}^{(a)}_{1}-\tilde{\beta}^{(a)}_{1}+i$, 
$-2\tilde{\beta}^{(a)}_{1}$ 
(resp. 
$\tilde{z}^{(a)}_{1}+\tilde{\beta}^{(a)}_{1}-i$, 
$2 \tilde{\beta}^{(a)}_{1}$).
Now we shall take the Trotter limit $N \to \infty$ in (\ref{nlie1}). 
\begin{eqnarray}
&& \hspace{-20pt}
\mathcal{T}^{(a)}_{m}(v)=Q^{(a)}_{m} \nonumber \\ 
&& +
\oint_{C^{(a)}_{m}} \frac{{\mathrm d} y}{2\pi i} 
\frac{1}{v-y-\beta^{(a)}_{m}}
 \bigg\{ 
 \frac{\mathcal{T}^{(a)}_{m-1}(y+\beta^{(a)}_{m}-\frac{i}{2}) 
 \mathcal{T}^{(a)}_{m+1}(y+\beta^{(a)}_{m}-\frac{i}{2})}
 {\mathcal{T}^{(a)}_{m}(y+\beta^{(a)}_{m}-i)} \nonumber \\
 && \hspace{90pt} +
 \frac{\mathcal{T}^{(a-1)}_{m}(y+\beta^{(a)}_{m}-\frac{i}{2}) 
 \mathcal{T}^{(a+1)}_{m}(y+\beta^{(a)}_{m}-\frac{i}{2})}
 {\mathcal{T}^{(a)}_{m}(y+\beta^{(a)}_{m}-i)}
 \bigg\} \nonumber \\
&& +
\oint_{\overline{C}^{(a)}_{m}} \frac{{\mathrm d} y}{2\pi i} 
 \frac{1}
 {v-y+\beta^{(a)}_{m}}
 \bigg\{ 
 \frac{\mathcal{T}^{(a)}_{m-1}(y-\beta^{(a)}_{m}+\frac{i}{2}) 
 \mathcal{T}^{(a)}_{m+1}(y-\beta^{(a)}_{m}+\frac{i}{2})}
 {\mathcal{T}^{(a)}_{m}(y-\beta^{(a)}_{m}+i)} \nonumber \\
 && \hspace{90pt} +
 \frac{\mathcal{T}^{(a-1)}_{m}(y-\beta^{(a)}_{m}+\frac{i}{2}) 
 \mathcal{T}^{(a+1)}_{m}(y-\beta^{(a)}_{m}+\frac{i}{2})}
 {\mathcal{T}^{(a)}_{m}(y-\beta^{(a)}_{m}+i)}
 \bigg\}
 \nonumber \\ 
 && \hspace{100pt} 
 {\rm for} \quad a \in \{1,2,\dots,r \}
 \quad {\rm and} \quad m \in {\mathbb Z}_{\ge 1},
 \label{nlie3}
\end{eqnarray}
where $\mathcal{T}^{(a)}_{m}(v):=
\lim_{N \to \infty} \widetilde{T}^{(a)}_{m}(v)$; 
$\beta^{(a)}_{m}:=\lim_{N \to \infty} \tilde{\beta}^{(a)}_{m}
=\frac{m+a}{2}i$; 
$\mathcal{T}^{(r+1)}_{m}(v)=
e^{\frac{m(\mu_{1}+\cdots +\mu_{r+1})}{T}}$; 
$\mathcal{T}^{(a)}_{0}(v)=1$;  
$\mathcal{T}^{(0)}_{m}(v)=
        \exp \left(\frac{mJ}{(v^2+\frac{m^2}{4})T}\right)$; 
 the contour $C^{(a)}_{m}$ (resp. $\overline{C}^{(a)}_{m}$) 
is a counterclockwise closed loop around $0$ 
 which satisfies the condition 
$|y| < |v-\beta^{(a)}_{m}|$ 
(resp. $|y| < |v+\beta^{(a)}_{m}|$)
and does not surround 
$-2\beta^{(a)}_{m}$ (resp. $2\beta^{(a)}_{m}$). 
In particular for $m=1$, we obtain a system of NLIE, which
contains only a {\em finite} number of unknown 
functions $\{{\mathcal T}^{(a)}_{1}(v) \}_{1\le a \le r}$: 
\begin{eqnarray}
{\mathcal T}^{(a)}_{1}(v)=Q^{(a)}_{1} 
&+&
\oint_{C^{(a)}_{1}} \frac{{\mathrm d} y}{2\pi i} 
 \frac{\mathcal{T}^{(a-1)}_{1}(y+\beta^{(a)}_{1}-\frac{i}{2}) 
 \mathcal{T}^{(a+1)}_{1}(y+\beta^{(a)}_{1}-\frac{i}{2})}
 {(v-y-\beta^{(a)}_{1})\mathcal{T}^{(a)}_{1}(y+\beta^{(a)}_{1}-i)}
 \nonumber \\
&+&
\oint_{\overline{C}^{(a)}_{1}} \frac{{\mathrm d} y}{2\pi i} 
 \frac{\mathcal{T}^{(a-1)}_{1}(y-\beta^{(a)}_{1}+\frac{i}{2}) 
 \mathcal{T}^{(a+1)}_{1}(y-\beta^{(a)}_{1}+\frac{i}{2})}
 {(v-y+\beta^{(a)}_{1})\mathcal{T}^{(a)}_{1}(y-\beta^{(a)}_{1}+i)}
 \nonumber \\ 
 && \hspace{120pt} 
 {\rm for} \quad a \in \{1,2,\dots,r\},
 \label{nlie4}
\end{eqnarray}
where $\mathcal{T}^{(r+1)}_{1}(v)=e^{\frac{\mu_{1}+\cdots +\mu_{r+1}}{T}}$; 
the contour $C^{(a)}_{1}$ (resp. $\overline{C}^{(a)}_{1}$) 
is a counterclockwise closed loop around $0$ 
which satisfies the condition 
$|y| < |v-\beta^{(a)}_{1}|$ 
(resp. $|y| < |v+\beta^{(a)}_{1}|$) and 
does not surround 
$z^{(a)}_{1}-\beta^{(a)}_{1}+i$, 
$-2\beta^{(a)}_{1}$ 
(resp. 
$z^{(a)}_{1}+\beta^{(a)}_{1}-i$, 
$2 \beta^{(a)}_{1}$); 
$z^{(a)}_{1}=\lim_{N\to \infty }{\tilde z}^{(a)}_{1}$.
One can calculate the free energy per site $f$ 
by using (\ref{nlie4}) 
and the relation
\begin{eqnarray}
f=J-T\log \mathcal{T}^{(1)}_{1}(0). 
\label{free-en}
\end{eqnarray}
We need not use (\ref{nlie3}) 
(for $m \in {\mathbb Z}_{\ge 2}$) to obtain the 
free energy per site. 
In addition, 
$\{{\mathcal T}^{(a)}_{m}(v) \}_{1\le a \le r;m\in{\mathbb Z}_{\ge 1}}$ 
 is given in terms of the solution of (\ref{nlie4}) 
$\{{\mathcal T}^{(a)}_{1}(v) \}_{1\le a \le r}$ 
based on a Jacobi-Trudi formula \cite{BR90}. 
However, (\ref{nlie3}) 
has theoretical significance since it 
 manifests the relation between our new NLIE and the 
 traditional TBA equations 
 (\ref{TBA-1}), (\ref{TBA-2}), (\ref{TBA-3}). 
In fact, (\ref{nlie3}) and 
(\ref{nlie4}) are related to the traditional 
TBA equations (\ref{TBA-1}), (\ref{TBA-2}), (\ref{TBA-3}) 
through the relation (\ref{Y-fun}) in the Trotter limit. 
\section{High temperature expansion}
\begin{figure}
\begin{center}
\includegraphics[width=0.95\textwidth]
{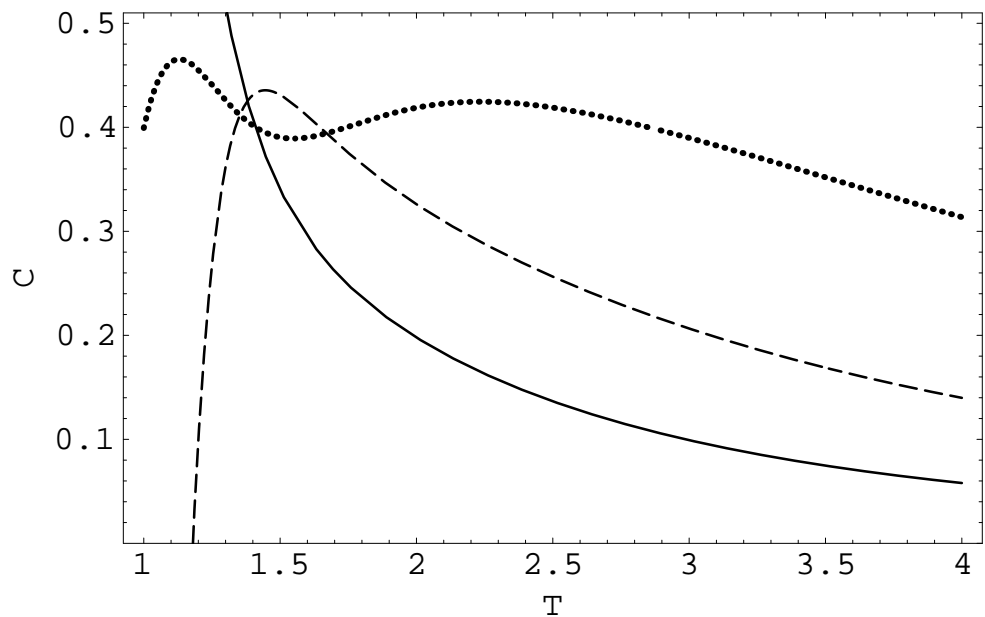}
\end{center}
\caption{Temperature dependence of the high temperature 
expansion of the specific heat of order 8 
for $sl(3)$, $J=1$, $\mu_{1}=h,\mu_{2}=0,\mu_{3}=-h$ case 
 ($h=0$: smooth line; $h=2$: broken line; $h=4$: dotted line.)
 }
\label{specific}
\end{figure}

In this section, we shall calculate the high temperature expansion of 
the free energy (\ref{free-en}) for $r=1,2,3$ 
by our new NLIE (\ref{nlie4}).  
As for the $XXX$-model case, 
the high temperature expansion of the free energy 
is calculated \cite{ShT02} by Takahashi's NLIE  
up to order 100. 
We assume the following expansion for large $T$: 
\begin{eqnarray}
\mathcal{T}^{(a)}_{1}(v)=
 \exp \left(\sum_{n=0}^{\infty}b_{n}^{(a)}(v)(\frac{J}{T})^{n} \right).
 \label{t-expan}
\end{eqnarray}
In contrasts with the $XXX$-model 
case \cite{ShT02}, 
we need further assumption 
\begin{eqnarray}
b_{n}^{(a)}(v)=\sum_{j=0}^{n-1}
 \frac{c_{n,j}^{(a)}v^{2j}}{(v^2+\frac{(a+1)^2}{4})^n},
\end{eqnarray}
where $c_{n,j}^{(a)} \in {\mathbb C}$.
Substituting (\ref{t-expan}) into (\ref{nlie4}), we 
can obtain the coefficients $\{b_{n}^{(a)}(v)\}$. 
Here we just enumerate first few of them. 
$b_{0}^{(a)}(v)$ has the form 
$b_{0}^{(a)}(v)=\log Q^{(a)}_{1}$. 
Other coefficients are given as follows. 
\\
$sl(2)$ case: 
\begin{eqnarray}
&& c^{(1)}_{1,0}=\frac{2\, Q^{(2)}_{1}}{{Q^{(1)}_{1}}^2} ,\nonumber \\
&& c^{(1)}_{2,0}=\frac{3\, Q^{(2)}_{
            1}}{{Q^{(1)}_{1}}^2} - \frac{6\, {Q^{(2)}_{1}}^2}{{Q^{(1)}_{
                1}}^4} ,\nonumber \\
&& c^{(1)}_{2,1}=\frac{Q^{(2)}_{1}}{{Q^{(1)}_{1}}^2} - 
\frac{4\, {Q^{(2)}_{1}}^2}{{Q^{(1)}_{
                1}}^4} ,\nonumber \\
&& c^{(1)}_{3,0}=\frac{10\, Q^{(2)}_{
            1}}{3\, {Q^{(1)}_{1}}^2} - \frac{18\, 
            {Q^{(2)}_{1}}^2}{{Q^{(1)}_{1}}^4} + 
            \frac{80\, {Q^{(2)}_{1}}^3}{3\, {Q^{(1)}_{1}}^6} ,\nonumber \\
&& c^{(1)}_{3,1}=\frac{3\, Q^{(2)}_{1}}{{Q^{(1)}_{1}}^2} - 
 \frac{22\, {Q^{(2)}_{1}}^2}{{Q^{(1)}_{1}}^4} + \frac{40\, {Q^{(2)}_{1}}^3}
 {{Q^{(1)}_{1}}^6} ,\nonumber \\
&& c^{(1)}_{3,2}=\frac{Q^{(2)}_{1}}{{Q^{(1)}_{1}}^2}
 - \frac{8\, {Q^{(2)}_{1}}^2}{{Q^{(1)}_{ 
                1}}^4} + \frac{16\, {Q^{(2)}_{1}}^3}{{Q^{(1)}_{ 
                1}}^6}. \label{sl2} 
\end{eqnarray}
$sl(3)$ case: 
\begin{eqnarray}
&& c^{(1)}_{1,0}=\frac{2\, Q^{(2)}_{1}}{{Q^{(1)}_{1}}^2} ,\nonumber \\
&& c^{(1)}_{2,0}=\frac{3\, Q^{(2)}_{ 
            1}}{{Q^{(1)}_{1}}^2} - \frac{6\, {Q^{(2)}_{1}}^2}{{Q^{(1)}_{1}}^4}
             + \frac{3\, 
        Q^{(3)}_{1}}{{Q^{(1)}_{1}}^3} ,\nonumber \\
&& c^{(1)}_{2,1}=\frac{Q^{(2)}_{1}}{{Q^{(1)}_{1}}^2}
 - \frac{4\, {Q^{(2)}_{1}}^2}{{Q^{(1)}_{1}}^4} + \frac{3\, 
        Q^{(3)}_{1}}{{Q^{(1)}_{1}}^3} ,\nonumber \\
&& c^{(1)}_{3,0}=\frac{10\, Q^{(2)}_{1}}{3\, {Q^{(1)}_{1}}^2} - 
\frac{18\, {Q^{(2)}_{1}}^2}{{Q^{(1)}_{1}}^4} + 
\frac{80\, {Q^{(2)}_{1}}^3}{3\, {Q^{(1)}_{1}}^6} + \frac{8\, 
        Q^{(3)}_{1}}{{Q^{(1)}_{1}}^3} - \frac{24\, Q^{(2)}_{1}\, 
        Q^{(3)}_{1}}{{Q^{(1)}_{1}}^5} ,\nonumber \\
&& c^{(1)}_{3,1}=\frac{3\, Q^{(2)}_{1}}{{Q^{(1)}_{1}}^2} - 
\frac{22\, {Q^{(2)}_{1}}^2}{{Q^{(1)}_{1}}^4} 
+ \frac{40\, {Q^{(2)}_{1}}^3}{{Q^{(1)}_{1}}^6} + \frac{13\, 
        Q^{(3)}_{1}}{{Q^{(1)}_{1}}^3} - \frac{42\, Q^{(2)}_{1}\, 
        Q^{(3)}_{1}}{{Q^{(1)}_{1}}^5} ,\nonumber \\
&& c^{(1)}_{3,2}=\frac{Q^{(2)}_{1}}{{Q^{(1)}_{1}}^2} - 
\frac{8\, {Q^{(2)}_{1}}^2}{{Q^{(1)}_{1}}^4} 
+ \frac{16\, {Q^{(2)}_{1}}^3}{{Q^{(1)}_{1}}^6} + \frac{5\, 
        Q^{(3)}_{1}}{{Q^{(1)}_{1}}^3} - \frac{18\, Q^{(2)}_{1}\, 
        Q^{(3)}_{1}}{{Q^{(1)}_{1}}^5}, \nonumber \\
\end{eqnarray}
\begin{eqnarray}
&& c^{(2)}_{1,0}=\frac{3\, Q^{(3)}_{1}}{Q^{(1)}_{1}\, Q^{(2)}_{1}} ,\nonumber \\
&& c^{(2)}_{2,0}=\frac{-27\, Q^{(3)}_{1}}{2\, {Q^{(1)}_{1}}^3} 
+ \frac{9\, Q^{(3)}_{1}}{Q^{(1)}_{1}\, 
        Q^{(2)}_{1}} - \frac{9\, {Q^{(3)}_{1}}^2}{2\, {Q^{(1)}_{1}}^2\, {Q^{(2)}_{
      1}}^2} ,\nonumber \\
&& c^{(2)}_{2,1}=\frac{-6\, Q^{(3)}_{1}}{{Q^{(1)}_{1}}^3} 
+ \frac{2\, Q^{(3)}_{1}}{Q^{(1)}_{1}\, 
        Q^{(2)}_{1}} ,\nonumber \\
&& c^{(2)}_{3
,0}=\frac{-1377\, Q^{(3)}_{1}}{16\, {Q^{(1)}_{1}}^3} + 
\frac{171\, Q^{(3)}_{1}}{8\, Q^{(1)}_{1}\, 
        Q^{(2)}_{1}} + \frac{243\, Q^{(2)}_{1}\, 
        Q^{(3)}_{1}}{2\, {Q^{(1)}_{1}}^5} 
        - \frac{27\, {Q^{(3)}_{
                1}}^2}{{Q^{(1)}_{1}}^2\, {Q^{(2)}_{1}}^2} 
     \nonumber \\
&& \hspace{30pt} - \frac{81\, {Q^{(3)}_{
                1}}^2}{16\, {Q^{(1)}_{1}}^4\, 
        Q^{(2)}_{1}} + \frac{9\, {Q^{(3)}_{1}}^3}{{Q^{(1)}_{1}}^3\, {Q^{(2)}_{
                1}}^3} ,\nonumber \\
&& c^{(2)}_{3
,1}=\frac{-135\, Q^{(3)}_{1}}{2\, {Q^{(1)}_{1}}^3}
 + \frac{12\, Q^{(3)}_{1}}{Q^{(1)}_{1}\, 
        Q^{(2)}_{1}} + \frac{108\, Q^{(2)}_{1}\, 
        Q^{(3)}_{1}}{{Q^{(1)}_{1}}^5} - \frac{6\, {Q^{(3)}_{1}}^2}{{Q^{(1)}_{1}}^2\, 
        {Q^{(2)}_{1}}^2} - \frac{45\, {Q^{(3)}_{1}}^2}{2\, {Q^{(1)}_{1}}^4\, 
        Q^{(2)}_{1}} ,\nonumber \\
&& c^{(2)}_{3
,2}=\frac{-13\, Q^{(3)}_{1}}{{Q^{(1)}_{1}}^3} + \frac{2\, Q^{(3)}_{1}}{Q^{(1)}_{1}\, 
        Q^{(2)}_{1}} + \frac{24\, Q^{(2)}_{1}\, 
        Q^{(3)}_{1}}{{Q^{(1)}_{1}}^5} - \frac{9\, {Q^{(3)}_{1}}^2}{{Q^{(1)}_{1}}^4\, 
        Q^{(2)}_{1}}. \nonumber \\
\end{eqnarray}
$sl(4)$ case: 
\begin{eqnarray}
&& c^{(1)}_{1
,0}=\frac{2\, Q^{(2)}_{1}}{{Q^{(1)}_{1}}^2} ,\nonumber \\
&& c^{(1)}_{2
,0}=\frac{3\, Q^{(2)}_{
            1}}{{Q^{(1)}_{1}}^2} - \frac{6\, {Q^{(2)}_{1}}^2}{{Q^{(1)}_{
                1}}^4} + \frac{3\, 
        Q^{(3)}_{1}}{{Q^{(1)}_{1}}^3} ,\nonumber \\
&& c^{(1)}_{2
,1}=\frac{Q^{(2)}_{1}}{{Q^{(1)}_{1}}^2}
 - \frac{4\, {Q^{(2)}_{1}}^2}{{Q^{(1)}_{1}}^4} + \frac{3\, 
        Q^{(3)}_{1}}{{Q^{(1)}_{1}}^3} ,\nonumber \\
&& c^{(1)}_{3
,0}=\frac{10\, Q^{(2)}_{
            1}}{3\, {Q^{(1)}_{1}}^2} - \frac{18\, {Q^{(2)}_{1}}^2}{{Q^{(1)}_{
                1}}^4} + \frac{80\, {Q^{(2)}_{1}}^3}{3\, {Q^{(1)}_{1}}^6}
                 + \frac{8\, 
        Q^{(3)}_{1}}{{Q^{(1)}_{1}}^3} - \frac{24\, Q^{(2)}_{1}\, 
        Q^{(3)}_{1}}{{Q^{(1)}_{1}}^5} + \frac{4\, 
        Q^{(4)}_{1}}{{Q^{(1)}_{1}}^4} ,\nonumber \\
&& c^{(1)}_{3
,1}=\frac{3\, Q^{(2)}_{
            1}}{{Q^{(1)}_{1}}^2} - \frac{22\, {Q^{(2)}_{1}}^2}{{Q^{(1)}_{
                1}}^4} + \frac{40\, {Q^{(2)}_{1}}^3}{{Q^{(1)}_{1}}^6}
                 + \frac{13\, 
        Q^{(3)}_{1}}{{Q^{(1)}_{1}}^3} - \frac{42\, Q^{(2)}_{1}\, 
        Q^{(3)}_{1}}{{Q^{(1)}_{1}}^5} + \frac{8\, 
        Q^{(4)}_{1}}{{Q^{(1)}_{1}}^4} ,\nonumber \\
&& c^{(1)}_{3
,2}=\frac{Q^{(2)}_{1}}{{Q^{(1)}_{1}}^2} - \frac{8\, {Q^{(2)}_{1}}^2}{{Q^{(1)}_{
                1}}^4} + \frac{16\, {Q^{(2)}_{1}}^3}{{Q^{(1)}_{1}}^6} + \frac{5\, 
        Q^{(3)}_{1}}{{Q^{(1)}_{1}}^3} - \frac{18\, Q^{(2)}_{1}\, 
        Q^{(3)}_{1}}{{Q^{(1)}_{1}}^5} + \frac{4\, 
        Q^{(4)}_{1}}{{Q^{(1)}_{1}}^4}, \nonumber \\
\end{eqnarray}
\begin{eqnarray}
&& c^{(2)}_{1
,0}=\frac{3\, Q^{(3)}_{1}}{Q^{(1)}_{1}\, Q^{(2)}_{1}} ,\nonumber \\
&& c^{(2)}_{2,0}=\frac{-27\, Q^{(3)}_{1}}{2\, {Q^{(1)}_{1}}^3} 
+ \frac{9\, Q^{(3)}_{1}}{Q^{(1)}_{1}\, 
        Q^{(2)}_{1}} - \frac{9\, {Q^{(3)}_{1}}^2}{2\, {Q^{(1)}_{1}}^2\, {Q^{(2)}_{
                1}}^2} + \frac{9\, Q^{(4)}_{1}}{{Q^{(1)}_{1}}^2\, 
        Q^{(2)}_{1}} ,\nonumber \\
&& c^{(2)}_{2
,1}=\frac{-6\, Q^{(3)}_{1}}{{Q^{(1)}_{1}}^3} 
+ \frac{2\, Q^{(3)}_{1}}{Q^{(1)}_{1}\, 
        Q^{(2)}_{1}} + \frac{4\, Q^{(4)}_{1}}{{Q^{(1)}_{1}}^2\, 
        Q^{(2)}_{1}} ,\nonumber \\
&& c^{(2)}_{3,0}=\frac{-1377\, Q^{(3)}_{1}}{16\, {Q^{(1)}_{1}}^3} 
+ \frac{171\, Q^{(3)}_{1}}{8\, Q^{(1)}_{1}\, Q^{(2)}_{1}} 
 + \frac{243\, Q^{(2)}_{1}\, Q^{(3)}_{1}}{2\, {Q^{(1)}_{1}}^5} 
 - \frac{27\, {Q^{(3)}_{1}}^2}{{Q^{(1)}_{1}}^2\, {Q^{(2)}_{1}}^2} 
 - \frac{81\, {Q^{(3)}_{1}}^2}{16\, {Q^{(1)}_{1}}^4\, Q^{(2)}_{1}} 
 \nonumber \\ 
   && \hspace{30pt}
    + \frac{9\, {Q^{(3)}_{1}}^3}{{Q^{(1)}_{1}}^3\, {Q^{(2)}_{1}}^3} 
 - \frac{81\, Q^{(4)}_{1}}{{Q^{(1)}_{1}}^4} 
 + \frac{783\, 
        Q^{(4)}_{1}}{16\, {Q^{(1)}_{1}}^2\, Q^{(2)}_{1}}
         - \frac{27\, Q^{(3)}_{1}\, 
        Q^{(4)}_{1}}{{Q^{(1)}_{1}}^3\, {Q^{(2)}_{1}}^2} ,\nonumber \\
&& c^{(2)}_{3,1}=\frac{-135\, Q^{(3)}_{1}}{2\, {Q^{(1)}_{1}}^3} 
+ \frac{12\, Q^{(3)}_{1}}{Q^{(1)}_{1}\,Q^{(2)}_{1}} 
+ \frac{108\, Q^{(2)}_{1}\,  Q^{(3)}_{1}}{{Q^{(1)}_{1}}^5} 
-  \frac{6\, {Q^{(3)}_{1}}^2}{{Q^{(1)}_{1}}^2\, {Q^{(2)}_{1}}^2}
\nonumber \\
&& \hspace{30pt}
 - \frac{45\, {Q^{(3)}_{1}}^2}{2\, {Q^{(1)}_{1}}^4\, Q^{(2)}_{1}}
- \frac{72\, Q^{(4)}_{1}}{{Q^{(1)}_{1}}^4} 
+ \frac{75\, Q^{(4)}_{1}}{2\, {Q^{(1)}_{1}}^2\, Q^{(2)}_{1}} 
- \frac{12\, Q^{(3)}_{1}\, Q^{(4)}_{1}}{{Q^{(1)}_{1}}^3\, {Q^{(2)}_{1}}^2}
 ,\nonumber \\
&& c^{(2)}_{3
,2}=\frac{-13\, Q^{(3)}_{1}}{{Q^{(1)}_{1}}^3} + \frac{2\, Q^{(3)}_{1}}{Q^{(1)}_{1}\, 
        Q^{(2)}_{1}} + \frac{24\, Q^{(2)}_{1}\, 
        Q^{(3)}_{1}}{{Q^{(1)}_{1}}^5} - \frac{9\, {Q^{(3)}_{1}}^2}{{Q^{(1)}_{1}}^4\, 
        Q^{(2)}_{1}} - \frac{16\, Q^{(4)}_{1}}{{Q^{(1)}_{1}}^4} + \frac{7\, 
        Q^{(4)}_{1}}{{Q^{(1)}_{1}}^2\, Q^{(2)}_{1}}, \nonumber \\
\end{eqnarray}
\begin{eqnarray}
&& c^{(3)}_{1
,0}=\frac{4\, Q^{(4)}_{1}}{Q^{(1)}_{1}\, Q^{(3)}_{1}} ,\nonumber \\
&& c^{(3)}_{2
,0}=\frac{20\, Q^{(4)}_{1}}{Q^{(1)}_{1}\, Q^{(3)}_{1}} - \frac{32\, Q^{(2)}_{1}\, 
        Q^{(4)}_{1}}{{Q^{(1)}_{1}}^3\, 
        Q^{(3)}_{1}} - \frac{8\, {Q^{(4)}_{1}}^2}{{Q^{(1)}_{1}}^2\, {Q^{(3)}_{
                1}}^2} ,\nonumber \\
&& c^{(3)}_{2
,1}=\frac{3\, Q^{(4)}_{1}}{Q^{(1)}_{1}\, Q^{(3)}_{1}} - \frac{8\, Q^{(2)}_{1}\, 
        Q^{(4)}_{1}}{{Q^{(1)}_{1}}^3\, Q^{(3)}_{1}} ,\nonumber \\
&& c^{(3)}_{3,0}=\frac{-192\, Q^{(4)}_{1}}{{Q^{(1)}_{1}}^4} 
+ \frac{248\, Q^{(4)}_{1}}{3\, Q^{(1)}_{1}\,Q^{(3)}_{1}} 
- \frac{352\, Q^{(2)}_{1}\, Q^{(4)}_{1}}{{Q^{(1)}_{1}}^3\, Q^{(3)}_{1}}
 + \frac{512\, {Q^{(2)}_{1}}^2\, Q^{(4)}_{1}}{{Q^{(1)}_{1}}^5\, Q^{(3)}_{1}}
 \nonumber \\ 
 && \hspace{30pt}
  - \frac{80\, {Q^{(4)}_{1}}^2}{{Q^{(1)}_{1}}^2\, {Q^{(3)}_{1}}^2} 
  + \frac{128\, Q^{(2)}_{1}\, {Q^{(4)}_{1}}^2}{{Q^{(1)}_{1}}^4\, {Q^{(3)}_{1}}^2}
   + \frac{64\, {Q^{(4)}_{1}}^3}{3\, {Q^{(1)}_{1}}^3\, {Q^{(3)}_{1}}^3} ,\nonumber \\
&& c^{(3)}_{3,1}=\frac{-96\, Q^{(4)}_{1}}{{Q^{(1)}_{1}}^4} 
+ \frac{30\, Q^{(4)}_{1}}{Q^{(1)}_{1}\, Q^{(3)}_{1}} 
- \frac{160\, Q^{(2)}_{1}\, Q^{(4)}_{1}}{{Q^{(1)}_{1}}^3\, Q^{(3)}_{1}} 
+ \frac{256\, {Q^{(2)}_{1}}^2\, Q^{(4)}_{1}}{{Q^{(1)}_{1}}^5\, Q^{(3)}_{1}} 
\nonumber \\
&& \hspace{30pt}
- \frac{12\, {Q^{(4)}_{1}}^2}{{Q^{(1)}_{1}}^2\, {Q^{(3)}_{1}}^2}
 + \frac{32\,Q^{(2)}_{1}\, {Q^{(4)}_{1}}^2}{{Q^{(1)}_{1}}^4\, {Q^{(3)}_{1}}^2} 
 ,\nonumber \\
&& c^{(3)}_{3
,2}=\frac{-12\, Q^{(4)}_{1}}{{Q^{(1)}_{1}}^4} + \frac{3\, Q^{(4)}_{1}}{Q^{(1)}_{1}\, 
        Q^{(3)}_{1}} - \frac{18\, Q^{(2)}_{1}\, Q^{(4)}_{1}}{{Q^{(1)}_{1}}^3\, 
        Q^{(3)}_{1}} + \frac{32\, {Q^{(2)}_{1}}^2\, Q^{(4)}_{1}}{{Q^{(1)}_{1}}^5\, 
        Q^{(3)}_{1}}. \nonumber \\
\end{eqnarray}
When $Q^{(a)}_{1}$ are suitably chosen, 
(\ref{sl2}) recover the known results\cite{ShT02}. 
One see that 
these coefficients are expressed in terms of the 
solutions of the $Q$-system (\ref{Q-sys}). 
And thus  a solution of the $T$-system 
is given in terms of solutions of the $Q$-system. 
The $T$-system is a Yang-Baxterization of the 
$Q$-system. 
Thus one may say that the degree of the expansion expresses the 
degree of a Yang-Baxterization. 
Using $\{b_{n}^{(a)}(0)\}$, we can calculate the 
free energy (\ref{free-en}) and the 
specific heat $C=-T\frac{\partial^{2} f}{\partial T^{2}} $.
We have plotted the high temperature expansion of the 
specific heat for $sl(3)$ in figure \ref{specific}. 
For large $T$, this agrees with the 
result from another NLIE  by Fujii and Kl\"umper 
(Fig.4 in  \cite{FK99}). 
This indicates the validity of our new NLIE (\ref{nlie4}).
We note that the high temperature expansion for the $SU(n)$ 
Heisenberg model is briefly reported in the letter \cite{FK02} 
based on a completely different method. 
%
\section{Concluding remarks}
In this paper, we have derived a system of NLIE 
with a {\em finite} number of unknown functions, 
which describes thermodynamics of the $sl(r+1)$ 
Uimin-Sutherland model. This type of NLIE 
 for  $sl(r+1)$ of {\em arbitrary} rank $r$ 
 is derived for the first time. 
 In particular for $r=1$, 
 our new NLIE (\ref{nlie4}) reduces to Takahashi's 
 NLIE \cite{Ta01} for the $XXX$ spin chain. 
 A relation between 
our new NLIE (\ref{nlie3}), (\ref{nlie4}) 
and the traditional TBA equations (\ref{TBA-1}),(\ref{TBA-2}), 
(\ref{TBA-3})  
which are also derived from the $T$-system is clarified. 
The high temperature expansion of the free energy 
is discussed, in which a solution of the $T$-system 
is given in terms of solutions of the $Q$-system. 
We expect that we can extend these results to other algebras 
 by using the $T$-systems in 
\cite{KNS95,KS95-2,T97,T98,T98-2,T99-1,T99}. 

In \cite{KW02}, 
 fugacity expansion formulae of the free energy for the 
 $XXX$ spin chain up to an infinite order 
is derived from the string center equation, 
 and Takahashi's  NLIE \cite{Ta01} is rederived from this formulae. 
 This work may be viewed in part as a kind of 
 Yang-Baxterization of the results in \cite{KN00} where 
  power series formulae related to a formal completeness of 
  the Bethe ansatz are 
derived from the string center equation. 
 We can also generally recover \cite{KNT02} 
 the results in \cite{KN00} from 
 the $Q$-system. 
Starting from our new NLIE (\ref{nlie4}), 
we may derive expansion formulae similar to the 
ones in \cite{KW02} for $sl(r+1)$ of arbitrary rank $r$, 
from which an idea toward a 
 Yang-Baxterization of the results in \cite{KNT02} may 
 be given since 
 our new NLIE is based on the $T$-system which 
 is a Yang-Baxterization of the $Q$-system. 

There are different types of 
NLIE with finite numbers of unknown functions 
for algebras of arbitrary rank 
 in rather different contexts \cite{Z98,DDT00}. 
 Their origin are related to Destri and de Vega's NLIE \cite{DD92}. 
 To derive analogous NLIE from the QTM method and 
to study the relation to our new NLIE deserve investigation. 
\section*{Acknowledgments}
\noindent
The author is financially supported by 
Inoue Foundation for Science. 
  
\end{document}